\documentclass{aastex6}
\usepackage{natbib}
\usepackage{epsfig,amsmath} 
\usepackage{graphicx}
\usepackage{amssymb}

\slugcomment{Received; published }
\journalinfo{Submitted to The Astrophysical Journal}

\newcommand{\be}{\begin{equation}}
\newcommand{\ee}{\end{equation}}
\newcommand{\bea}{\begin{eqnarray}}
\newcommand{\eea}{\end{eqnarray}}

\newcommand{\kappaR}{\kappa_{\rm R}}

\newcommand{\rmd}{{d}}
\newcommand{\fourx}{\underline{x}}
\newcommand{\fourp}{\underline{p}}
\newcommand{\vecx}{\mbox{\boldmath $x$}}
\newcommand{\vecp}{\mbox{\boldmath $p$}}
\newcommand{\vecF}{\mbox{\boldmath $F$}}
\newcommand{\vech}{\mbox{\boldmath $h$}}
\newcommand{\vn}{\mbox{\boldmath $\hat{n}$}}
\newcommand{\vomega}{\mbox{\boldmath $\hat{\omega}$}}
\newcommand{\vOmega}{\mbox{\boldmath $\hat{\Omega}$}}

\newcommand{\unb}{\underline{\nabla}}
\newcommand{\vnabla}{\mbox{\boldmath $\nabla$}}
\newcommand{\vnablat}{\mbox{\boldmath $\nabla_{\tau}$}}
\newcommand{\partialt}{\partial_{\tau}}
\newcommand{\noccx}{n} 

\newcommand{\noccpm}{\tilde{n}_{\pm}}
\newcommand{\noccp}{\tilde{n}_{+}}
\newcommand{\noccm}{\tilde{n}_{-}}
\newcommand{\re}{r_{\rm e}}

\newcommand{\mue}{\mu_{\rm e}}
\newcommand{\me}{m_{\rm e}}
\newcommand{\mprot}{m_{\rm p}}
\newcommand{\Ne}{N_{\rm e}}
\newcommand{\lambdac}{\lambda_{\rm C}}
\newcommand{\sigmat}{\sigma_{\rm T}}

\def\tens#1{\ensuremath{\mathsf{#1}}}
    
\begin{document}

% : spectra and color corrections. 
%Radiation pressure and  
 \title{Rosseland and flux mean opacities  for Compton scattering }

\shorttitle{Rosseland mean opacity  for Compton scattering}
\shortauthors{Juri Poutanen}

\author{
Juri Poutanen
}

\affil{Tuorla Observatory, Department of Physics and Astronomy, University of Turku, 
  V\"ais\"al\"antie 20, FI-21500 Piikki\"o, Finland; juri.poutanen@utu.fi} 
\affil{Nordita, KTH Royal Institute of Technology and Stockholm University, Roslagstullsbacken 23, SE-10691 Stockholm, Sweden
}

%\date  
 
\begin{abstract}
Rosseland mean opacity plays an important role in theories of stellar evolution and X-ray burst models. 
In the high-temperature regime, when most of the gas is completely ionized, the opacity is 
dominated by Compton scattering. 
Our aim here is to critically evaluate previous works on this subject and to compute 
exact Rosseland mean opacity  for Compton scattering 
in a broad range of temperatures and electron degeneracy parameter. 
We use relativistic kinetic equations for Compton scattering and 
compute the photon mean free path as a function of photon energy 
by solving the corresponding integral equation in the diffusion limit.   
As a byproduct we also demonstrate the way to compute photon redistribution functions in case of degenerate electrons. 
We then compute the  Rosseland mean opacity as a function of temperature and electron degeneracy.  
We compare our results to the previous calculations and find a significant difference in
the low-temperature regime and strong degeneracy. We find useful analytical expressions 
that approximate well the numerical results. 
We then proceed to compute the flux mean opacity and show that in diffusion approximation 
it is nearly identical  to the Rosseland mean opacity. We also provide a simple way
for accounting for the true absorption in evaluation of the Rosseland and flux mean opacities. 
\end{abstract}


%\maketitle
%
%________________________________________________________________

\section{Introduction}
\label{sec:intro}

The key role in the description of the radiation transport through the medium is played by two average opacities. 
The first one, known as the Rosseland mean opacity 
\be 
\kappa_{\rm R} = \int_0^\infty d\nu\ \left( \partial B_\nu/\partial T\right)  \left/ \int_0^\infty d\nu\ \kappa_{\nu}^{-1}  \left( \partial B_\nu /\partial T\right) ,
\right.     
\ee 
%$\kappa_{\rm R}$, 
relates the  temperature gradient to the radiation flux: 
\be 
\vecF= - \frac{ac}{3 \kappa_{\rm R}}   \vnabla{T^4}. 
\ee 
The second one, known as the flux mean opacity  
\be 
\kappa_{\rm F} =  \left. \int_0^\infty d\nu\ \kappa_\nu \ F_\nu   \right/ F ,  
\ee 
relates the bolometric radiation flux to the radiative acceleration \citep[see ][pp. 360-361]{Mihalas84}: 
\be 
\mbox{\boldmath $g$}_{\rm rad}=    \frac{ \kappa_{\rm F} }{c}   \vecF. 
\ee 
The Rosseland mean can be easily computed once the  total, absorption and scattering, opacity as a function of photon frequency $\kappa_{\nu}$ is known. 
For the flux mean, we also need to specify the spectral energy distribution  given by the flux $F_\nu$.
However,  in the diffusion approximation  these two opacities coincide for pure absorption and coherent scattering.

In the high-temperature regime, when most of the gas is completely ionized, the opacity is dominated by Compton scattering. 
This situation is not so simple as the scattering is incoherent, induced scattering has to be accounted for and instead of the total cross-section the effective cross-section should be used.
The case of the non-degenerate electron gas was  considered by \citet{Sampson:59}.
It was further extended by \citet{Chin65} to include the effect of electron degeneracy. 
This work was affected by an error, which also propagated to the textbooks \citep{Chiu68,CG68,Weiss04}. 
The corrected method to compute the Rosseland mean was introduced by \citet{BY76}, who provide also a comprehensive analysis of the previous results. 
%On the other hand, the limitations in computer power forced also \citet{BY76} to make approximation of coherent scattering at low temperatures. 
The numerical results presented in that work were approximated by \citet{Pacz:83} 
with a simple analytical expression, which were later used in numerous papers on X-ray bursts. 
An alternative approximation was given by \citet{WZW78}. 

In this paper we recompute the Rosseland and the flux mean opacities for  Compton scattering 
and  compare our results to the previous calculations. 
We also provide new analytical formulae that approximate  well the numerical results.

\section{Relativistic kinetic equation for Compton scattering}
\label{sec:rke}

Derivation of the Rosseland mean opacity for  Compton scattering 
is based on solution of the relativistic kinetic equation (RKE)  in terms of the photon 
mean-free path as a function of its energy. 
Interaction between photons and electrons (positrons) via Compton scattering accounting for the 
induced scattering and fermion degeneracy 
can be described by the explicitly covariant RKE for photons \citep{BY76,dGvLvW80,NP93,NP94}:
\bea \label{eq:rke}
\fourx \cdot \unb \noccx(\vecx) & = &  \frac{\re^2}{2}  \frac{2}{\lambdac^3}
\int \frac{\rmd \vecp}{\gamma} \frac{\rmd \vecp_1}{\gamma_1}  \frac{\rmd \vecx_1}{x_1} 
\: F  \: \delta^4(\fourp_1 + \fourx_1 - \fourp - \fourx)   \\ 
&\times & 
%\left\{ \noccx(\vecx_1) [1+\noccx(\vecx)]  \nocce(\vecp_1) [1-\nocce(\vecp)] \right. \\
\left\{ \noccx(\vecx_1) [1+\noccx(\vecx)]  \left[ \noccm(\vecp_1) (1\! -\!\noccm(\vecp)) + \noccp(\vecp_1) (1\! -\! \noccp(\vecp))  \right] \right. \nonumber \\
%&-& \left. \noccx(\vecx) [1+\noccx(\vecx_1)] \nocce(\vecp) [1-\nocce(\vecp_1)] ) \right\} , \nonumber
&-& \left. \noccx(\vecx) [1+\noccx(\vecx_1)]  \left[ \noccm(\vecp) (1\! -\! \noccm(\vecp_1)) + \noccp(\vecp) (1\! -\! \noccp(\vecp_1))  \right]    \right\} , \nonumber
\eea
where $\unb=\{\partial/c \partial t, -\vnabla\}$ is the four-gradient, 
$\re$ is the classical electron radius, $\lambdac=h/\me c$ is the Compton wavelength.   
Here we defined the dimensionless photon four-momentum as $\fourx =\{ x, \vecx \}= x \{ 1,\vomega\}$, where $\vomega$ 
is the unit vector in the photon propagation direction and $x \equiv h\nu/\me c^2$ is the photon energy in units of the electron rest mass. 
The photon distribution is described by the occupation number $\noccx$.
% or by the specific intensity (per dimensionless 
%energy interval) $I(\vecx)=x^3\noccx(\vecx)/C$, where constant $C$ is given by Eq.~(\ref{eq:const_int}).  
The dimensionless electron/positron four-momentum is $\fourp  = \{ \gamma, \vecp\}= \{ \gamma, p\vOmega\} = 
\gamma \{ 1 , \beta\vOmega\}$, 
where $\vOmega$ is the unit vector along the electron momentum, 
$\gamma$ and $p=\sqrt{\gamma^2-1}$  are the electron Lorentz factor and its momentum in units of $\me c$ and 
$\beta$ is the velocity in units of $c$.
The electron/positron distributions are described by the occupation numbers $\noccpm$. 
%If both electrons and positrons are present, equation  (\ref{eq:rke}) contains terms for both types of particles. 

The factor $F$ in Equation (\ref{eq:rke})  is the Klein--Nishina reaction rate \citep{LLVol4}
\begin{equation} \label{eq:kn}
F  = \left( \frac{1}{\xi} - \frac{1}{\xi_1}\right)^2 + 2 \; \left( \frac{1}{\xi} - \frac{1}{\xi_1}\right)
+ \frac{\xi}{\xi_1} + \frac{\xi_1}{\xi} , 
\end{equation}
and 
\begin{equation} \label{eq:xixi1}
\xi = \fourp_1\cdot\fourx_1= \fourp\cdot\fourx, \qquad \xi_1 =  \fourp_1\cdot\fourx =  \fourp\cdot\fourx_1
\end{equation}
are the four-products of the corresponding momenta. Second equalities in Eqs. (\ref{eq:xixi1}) arise  from the four-momentum conservation law represented 
by the delta-function in Eq.~(\ref{eq:rke}). 

The electron/positron distribution under assumption of thermal equilibrium and isotropy, 
is  given by the Fermi-Dirac distribution: 
\be \label{eq:fermi}
\noccpm(\vecp) = %  \frac{1}{ {\rm e}^{\displaystyle\frac{\gamma-1}{\Theta}-\eta}+1} . 
  \frac{1}{\displaystyle \exp\left(\frac{\gamma-1}{\Theta}-\eta_{\pm}\right)+1} , 
\ee
where $\Theta=kT /\me c^2$ is the dimensionless  temperature and $\eta_{\pm}$  are 
the degeneracy parameters for positron and electrons 
(the ratio of the Fermi energy minus rest mass to temperature) 
related via $\eta_- + \eta_+ = -2/\Theta$ (see e.g. \citealt{CG68}; page 302 of \citealt{Weiss04}).   
The electron/positron concentrations are given by the integrals over the momentum space:  
\be
N_\pm = 4\pi \frac{2}{\lambdac^3} \int_0^{\infty} p^2 \rmd p\ \noccpm(\vecp) , 
\ee
and  the density (not including electrons and positrons created by pair-production 
as well as radiation) is  
\be 
\rho= (N_- - N_+) \mue \mprot, 
\ee
where $\mue=2/(1+X)$ is the mean number of nucleons per free ionization electron 
and $X$ is the hydrogen mass fraction. The total number density of electrons and 
positrons is $\Ne= N_- + N_+$.
 
%In the limit $\eta\rightarrow-\infty$, electrons are non-degenerate and 

%For the isotropic electron distribution, we will use  the electron distribution function  
%to the electron density 
%$\fe(p)=2\nocce(\vecp)/\lambdac^3\Ne$, normalized to unity  
%\be
%4\pi \int_0^\infty \fe(p) \ p^2 \rmd p  = 1 . 
%\ee 

%In the following, let us consider a steady state 
%and ignore electron degeneracy, because in the upper atmosphere layers, 
%where the radiation spectrum is formed, electrons are non-degenerate. 
The form of the RKE  (\ref{eq:rke}) can be simplified by defining the redistribution 
functions (RF) via 
\be \label{eq:rf_gen}
R_\pm(\vecx_1 \rightarrow \vecx)  =  \frac{3}{16\pi}   \frac{2}{\lambdac^3}\frac{1}{N_\pm}
\int \! \frac{\rmd \vecp}{\gamma} \frac{\rmd \vecp_1}{\gamma_1}  \noccpm(\vecp_1) 
[1-\noccpm(\vecp)] \  
F\ \delta^4(\fourp_1 + \fourx_1 - \fourp - \fourx)  .
 \ee
%For the Maxwellian distribution of temperature $\Theta=k\Te /\me c^2$,
%\be \label{eq:maxwell}
%\fe(p) = \frac{1}{4\pi\ \Theta\ K_2(1/\Theta)} \exp(-\gamma/\Theta)
%\ee
%(where $K_2$ is the modified Bessel function),  
The RFs  satisfy the symmetry property  
\be\label{eq:rf_symm}
R_\pm(\vecx\rightarrow \vecx_1) \ e^{-x/\Theta} = R_\pm(\vecx_1 \rightarrow \vecx) \ e^{-x_1/\Theta} , 
%\exp\left( [x-x_1]/\Theta\right) ,
\ee
which follows from its definition (\ref{eq:rf_gen}) and the energy conservation $\gamma_1=\gamma+x-x_1$, 
or from the detailed balance  condition (see eq. 8.2 in \citealt{Pom73}).

In the absence of strong magnetic field, the medium is isotropic, therefore 
the RF depends only on the photon energies and the scattering angle (with $\mu$ being its cosine), 
i.e. we can write $R_\pm(\vecx_1\rightarrow \vecx) = R_\pm(x,x_1,\mu)$. 
%and the total scattering cross section (in units of the Thomson cross section $\sigmat$) as 
%\be \label{eq:totcross}
%\overline{s}_0(\vecx)  =\! 
%\frac{3}{16\pi}  \frac{1}{x} \! \int \!\! \frac{\rmd ^3 p}{\gamma} 
%\frac{\rmd ^3 p_1}{\gamma_{1}} \frac{\rmd ^3 x_1}{x_{1}}
% \fe(\gamma) \  F \ \delta (\fourp_1 + \fourx_1 - \fourp - \fourx)  .
%\ee
Introducing the total RF as 
\be\label{eq:rf_total}
R(x,x_1,\mu) = \frac{N_-}{\Ne} R_-(x,x_1,\mu)+\frac{N_+}{\Ne}  R_+(x,x_1,\mu) ,
\ee
the kinetic equation (\ref{eq:rke})  in a steady-state can be recast in a standard form of the radiative transfer equation
\be \label{eq:rte2}
%\lefteqn{ 
%\frac{1}{\sigmat \: \Ne}  
\vomega \cdot  \vnablat \noccx(\vecx) =
- \noccx(\vecx) \frac{1}{x}  \int_{0}^{\infty} \!\!\!\!  x_1 \rmd x_1 \!\! \int \!\!  
\rmd ^2 \vomega_1 \: R(x_1,x,\mu)\   [1+\noccx(\vecx_1)] 
+  [1+\noccx(\vecx)] \frac{1}{x}  
\int_{0}^{\infty} \!\!\!\!  x_1 \rmd x_1\!\! \int \!\! \rmd ^2 \vomega_1 \: R(x,x_1,\mu)\ \noccx(\vecx_1) ,
\ee
where $\vnablat=\vnabla/\sigmat\Ne$ is the dimensionless gradient, with $\sigmat$ being the Thomson cross-section.

\section{Photon mean free path} 
\label{sec:meanfreepath}

Deep inside stars or  thermonuclear burning regions of X-ray bursts, radiation field is nearly isotropic and the diffusion approximation should be rather accurate. 
We therefore can express the occupation number as 
\be\label{eq:n_diffuse}
\noccx(\vecx) = b_x - {l_x} \vomega\cdot \vnablat b_x ,
%\frac{l_x }{\sigmat \: \Ne}  
\ee
where $b_x=1/[\exp(x/\Theta)-1]$ is the occupation number for the Planck distribution and 
$l_x$ is the  mean free path (in units of $1/\sigmat\Ne$) for Compton scattering of a photon of energy $x$. 
Substituting expansion (\ref{eq:n_diffuse}) to Eq.~(\ref{eq:rte2}), 
noticing that the zeroth  order terms cancel out, keeping only terms of the first order in $\vnablat b_x$, 
and using condition (\ref{eq:rf_symm}), we  get \citep{Sampson:59,BY76}: 
\be\label{eq:diffuse}
\vomega \cdot \vnablat b_x \!= \!
\frac{1}{x}  \int_{0}^{\infty} \!\!\!\! x_1 \rmd x_1\!\! \int \! \rmd ^2 \vomega_1 \: R(x_1,x,\mu) 
\left[  l_x \vomega\cdot \vnablat b_x   \left(  \frac{1-e^{-x/\Theta}}{1-e^{-x_1/\Theta}}\right) 
- l_{x_1} \vomega_1 \cdot \vnablat b_{x_1} 
\left(  \frac{e^{x_1/\Theta}-1}{e^{x/\Theta}-1} \right) 
\right] . 
\ee
Simple algebra gives a linear integral equation for the mean free path $l_x$: 
\be\label{eq:diffuse2}
 1  =  \frac{1}{x}  \int_{0}^{\infty} \!\!\!\! x_1 \rmd x_1\!\! \int \! \rmd ^2 \vomega_1 \: R(x_1,x,\mu) 
% \left(  \right)
 \frac{1-e^{-x/\Theta}}{1-e^{-x_1/\Theta}} 
  \left[  l_x  - l_{x_1} \frac{x_1}{x} \frac{\vomega_1 \cdot \vnabla \Theta}{\vomega \cdot \vnabla \Theta} \right] . \nonumber
\ee
Choosing the coordinate system so that $\vomega=(0,0,1)$,  defining 
$\vomega_1=\left(\sqrt{1-\mu^2}\cos\phi,\sqrt{1-\mu^2}\sin\phi,\mu\right)$ and
$\nabla\Theta\propto (\sin\theta,0,\cos\theta)$, the integral over solid angle becomes $\int \rmd\mu \int \rmd \phi$, 
with only the last term in the  square brackets depending on $\phi$. The azimuthal integral is then 
\be
\int_0^{2\pi} \!\!\! \!\! \rmd \phi  \frac{\vomega_1 \cdot \vnabla \Theta}{\vomega \cdot \vnabla \Theta} =\!\!  
\int_0^{2\pi} \!\! \!\! \rmd \phi \ \left( \mu + \sqrt{1-\mu^2}\tan\theta \cos\phi \right) = 2\pi \mu, 
\ee
so that  the square bracket in Eq.~(\ref{eq:diffuse2}) can be substituted by $l_x  - l_{x_1} x_1 \mu/x$ \citep{Sampson:59}. 
Equation (\ref{eq:diffuse2}) can be further modified by integrating over the angles of the scattered photon: 
\bea\label{eq:integr_mean} 
1= 4\pi \int \limits_{0}^{\infty}  \frac{x_1}{x} \rmd x_1 
  \frac{1-e^{-x/\Theta}}{1-e^{-{x_1}/{\Theta}}} %  \\
%&\times& 
%\left[   R(x_1,x)   \left( l_x \!- \! l_{x_1}\! \frac{x_1}{x}\!  \right)\! + \! R^*\!(x_1,x)  l_{x_1}\!\!  \frac{x_1}{x} \! \right] , 
\left[ l_x R_0(x_1,x)    - l_{x_1}\! \frac{x_1}{x} R_1 (x_1,x)   \right], 
%\nonumber
 \eea
where we introduced the moments of the RF \citep{NP94}
\bea\label{eq:RF_moment_0}
 R_0(x_1,x) & = & \frac{1}{2} \int_{-1}^{1} R(x_1,x,\mu) \rmd \mu ,\\
\label{eq:RF_moment_1}
%  R^*(x_1,x) & = & \frac{1}{2} \int_{-1}^{1} R(x_1,x,\mu) \ (1-\mu)\ \rmd \mu. 
  R_1(x_1,x) & = & \frac{1}{2} \int_{-1}^{1} R(x_1,x,\mu) \ \mu\ \rmd \mu. 
\eea
The method for computing  these functions is described in Appendix \ref{sec:appa}.

%Equation (\ref{eq:integr_mean}) can be rewritten  as: 
%\bea\label{eq:cros_rel}
%l_x = \frac{ 1 + \overline{(x_1\mu/x)}\overline{s_0}(x) + \Delta(x)}{\overline{s'}(x)  }  
%l_x = \frac{ 1 + r_1(x) + \Delta_1 (x)}{ r_0(x)  + \Delta_0(x)  }  
%l_x = \frac{ 1 + r_1(x)  }{ r_0(x)    }  
%\eea
 
At low temperatures the RFs are extremely peaked at $x_1\approx x$ and 
therefore two approximations are often made \citep{Sampson:59,BY76}: 
\bea \label{eq:approx1}
l_{x_1} & \approx & l_x , \\ 
\label{eq:approx2}
\frac{1-e^{-x/\Theta}}{1-e^{-{x_1}/{\Theta}}}  & \approx & 1 . 
\eea
The first approximation is equivalent to the on-the-spot approximation in 
the theory of radiative transfer in spectral lines. 
These approximations reduce Eq.~(\ref{eq:integr_mean})  for the mean free path to 
\be \label{eq:lx_nonrel}
\frac{1}{l_x} \approx s_0(x)  - s_1(x) , 
%= \overline{(1 - x_1 \mu/x)}\ \overline{s_0}(x)   ,  
\ee
where \citep{NP94} 
\be
s_i(x)  =   \frac{4\pi}{x^{i+1}} \int x_1^{i+1} \rmd x_1     R_i (x_1,x)  , \quad i=0,1 . 
%s_1(x) & =&  \frac{4\pi}{x^2} \int x_1^2 \rmd x_1     R_1 (x_1,x)   . 
\ee

\begin{figure}
%\plotone{fig1.eps}
\centerline{\epsfig{file= 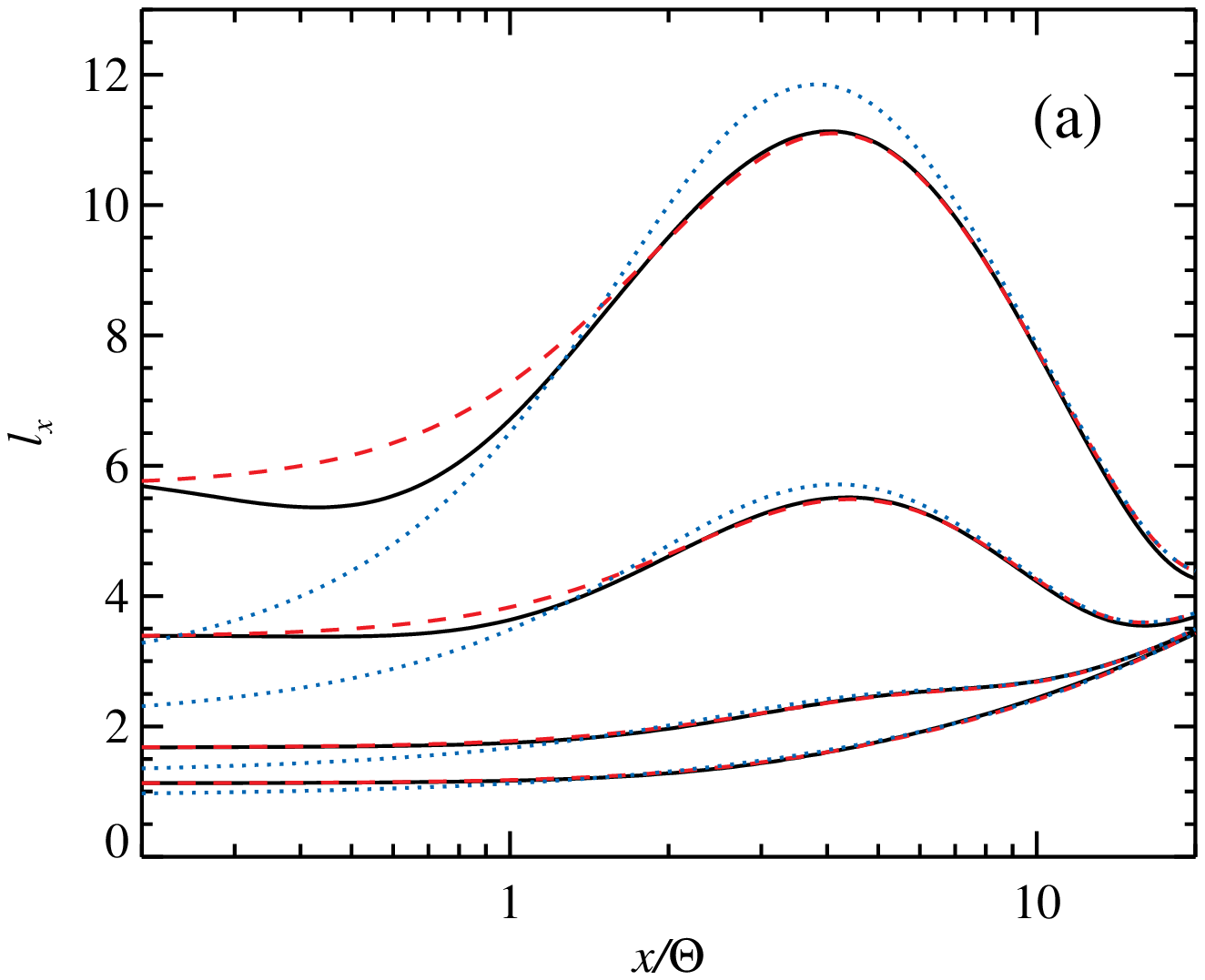,width=6.2cm}} 
\centerline{\epsfig{file= 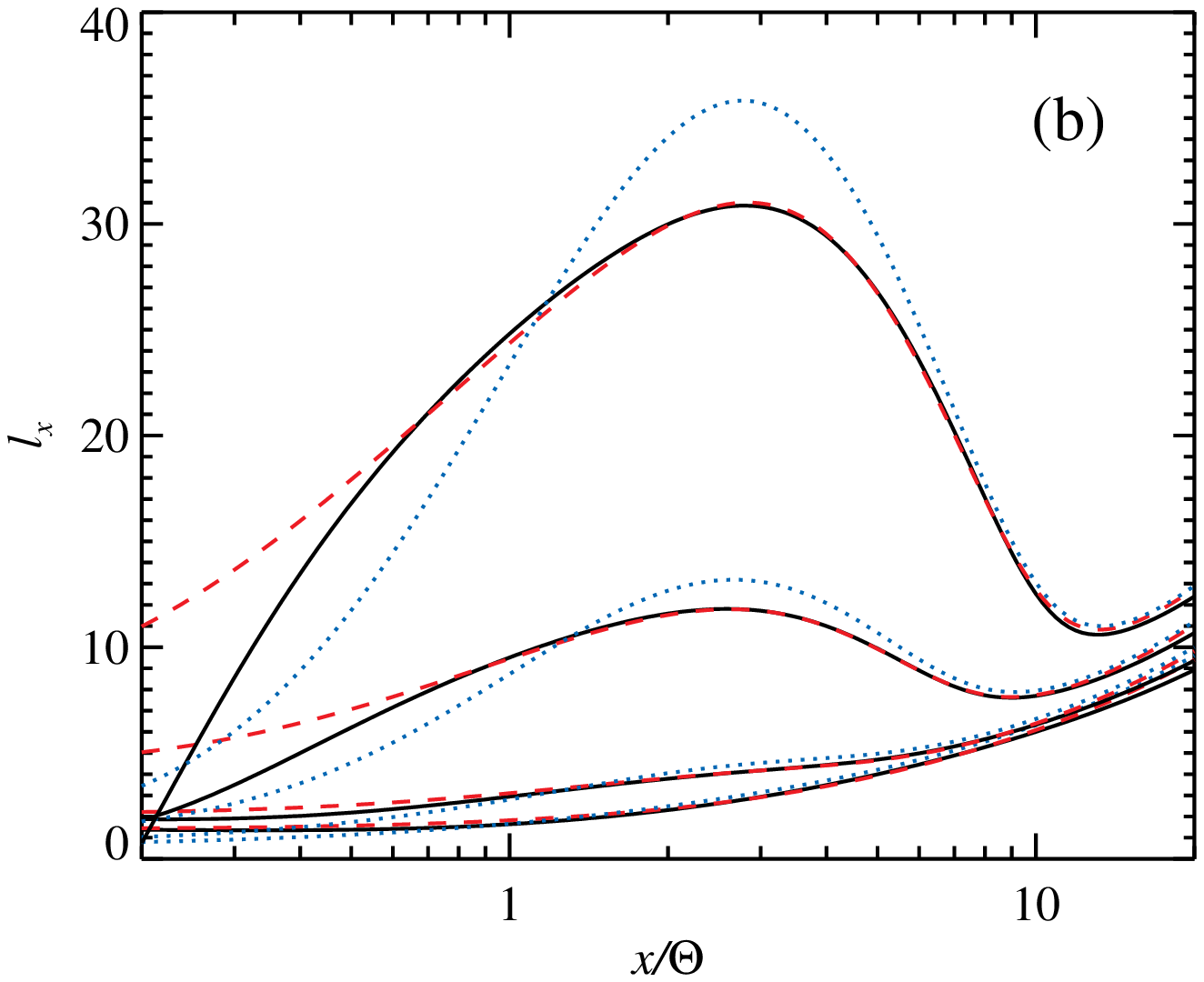,width=6.2cm}} 
\centerline{\epsfig{file= 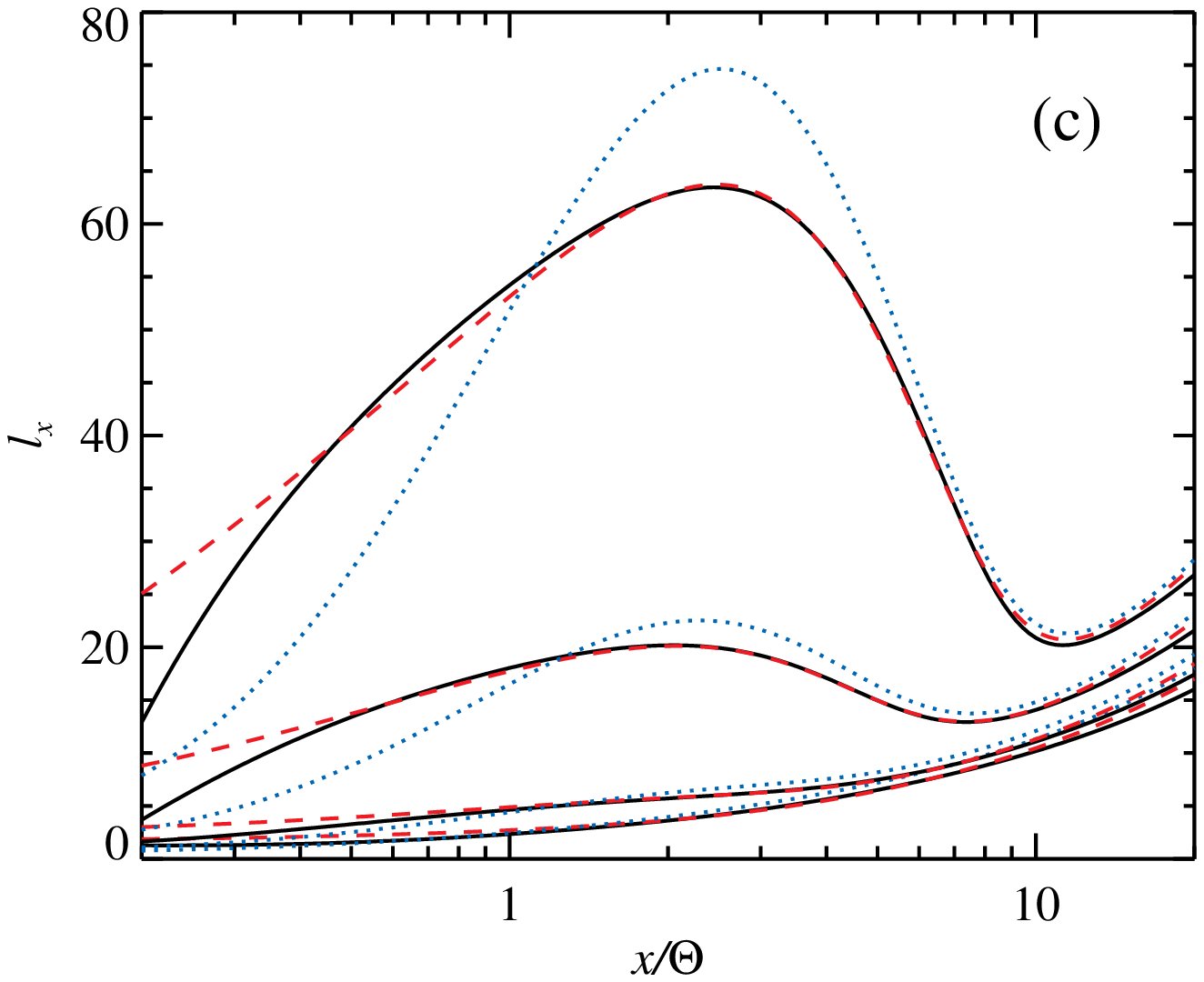,width=6.2cm}} 
\caption{Mean-free path of photons $l_x$ as a function of photon energy to temperature ratio 
for various temperatures and degeneracies: (a)  for $\Theta=0.05$, (b) $\Theta=0.25$, (c) $\Theta=0.5$. 
Different lines from bottom to the top correspond to the degeneracy parameter $\eta=-2, 1, 4, 7$. 
The exact solution (\ref{eq:integr_mean}) are shown by the  solid black lines. 
The approximate expressions (\ref{eq:lx_nonrel}) and (\ref{eq:lx_rel})  
are shown by the dotted blue and dashed red  lines, respectively. 
} 
\label{fig:meanfree}
\end{figure}

At  temperatures above 50 keV, approximation (\ref{eq:approx2}) fails (see Fig.~\ref{fig:meanfree}). 
Still keeping the on-the-spot approximation (\ref{eq:approx1}), we get an explicit expression 
%useful approximation 
\be  \label{eq:lx_rel}
\frac{1}{l_x} \approx r_0(x)  - r_1(x) ,
\ee
where 
%\be
%r_1^*(x) =  \frac{4\pi}{x^2} \int x_1^2 \rmd x_1     R_1 (x_1,x) \  \frac{1-e^{-x/\Theta}}{1-e^{-{x_1}/{\Theta}}}  . 
%\ee
\be
r_i(x)   =  \frac{4\pi}{x^{i+1}} \int x_1^{i+1} \rmd x_1     R_i (x_1,x)  \frac{1-e^{-x/\Theta}}{1-e^{-{x_1}/{\Theta}}}  .
%r_1(x) & =&  \frac{4\pi}{x^2} \int x_1^2 \rmd x_1     R_1 (x_1,x) \  l_{x_1}   \frac{1-e^{-x/\Theta}}{1-e^{-{x_1}/{\Theta}}}  . 
%r_0(x) &  =  & \frac{4\pi}{x} \int x_1 \rmd x_1     R_0 (x_1,x)   , \\
%\Delta_0(x) &  =  & 
%\frac{4\pi}{x} \int x_1 \rmd x_1     R_0 (x_1,x)  
%\left[ \frac{1-e^{-x/\Theta}}{1-e^{-{x_1}/{\Theta}}}   -1   \right]   , \\
%r_1(x) & =& \frac{4\pi}{x^2} \int x_1^2 \rmd x_1     R_1 (x_1,x)   , \\
%\Delta_1 (x) & = & \frac{4\pi}{x^2} \int x_1^2 \rmd x_1     R_1 (x_1,x)  
%\left[ l_{x_1}   \frac{1-e^{-x/\Theta}}{1-e^{-{x_1}/{\Theta}}}   -1   \right] . 
\ee
At low temperatures, the easiest way to exactly solve  Eq.~(\ref{eq:integr_mean})  for $l_x$ is 
to use iteration procedure, starting from the approximation (\ref{eq:lx_rel}). 
The functions $r_i(x)$  can be tabulated in advance. The integrals over 
the energy $x_1$ for every $x$ have to be taken over a dense grid around $x$.  
For high temperatures, in principle, one can replace the integral by the discrete sum on a logarithmic grid of photon energies $x_i$
and solve Eq.~(\ref{eq:integr_mean})  as a system of   linear equations  for $ l_i=l_{x_i}$ (as was done by \citealt{BY76}): 
\bea\label{eq:diffuse4}
\frac{1}{4\pi}  =  
 l_i\  a_i + \sum_j \ l_j\ b_{ij} = \sum_j \ l_j\ ( b_{ij}  + a_i \delta_{ij}) , 
% l_i \sum_j \! w_j  \frac{x_j^2}{x_i}   \frac{1-e^{-x_i/\Theta}}{1-e^{-{x_j}/{\Theta}}} R_0(x_j,x_i)  \nonumber \\
%&- &\sum_j \! w_j \frac{x_j^3}{x_i^2} \frac{1-e^{-x_i/\Theta}}{1-e^{-{x_j}/{\Theta}}} R_1(x_j,x_i)   l_j , 
%\nonumber
 \eea
 where 
\bea\label{eq:diffuse5}
a_i   & = & \sum_j \! w_j  \frac{x_j}{x_i}   \frac{1-e^{-x_i/\Theta}}{1-e^{-{x_j}/{\Theta}}} R_0(x_j,x_i)  , \\
b_{ij}   &=& - w_j \frac{x_j^2}{x_i^2} \frac{1-e^{-x_i/\Theta}}{1-e^{-{x_j}/{\Theta}}} R_1(x_j,x_i)   , 
%\nonumber
 \eea
and $w_j$ are the integration weights (equal to $x_j\ \Delta \ln x$ for a log-grid), 
$\delta_{ij}$ is the Kronecker delta. 
The results of calculations for $l_x$ using solution of the integral equation as well as by approximate formulae (\ref{eq:lx_rel}) and (\ref{eq:lx_nonrel}) 
are presented in Fig.~\ref{fig:meanfree}.

We see that the mean free path computed using expression  (\ref{eq:lx_rel}) approximates well the exact $l_x$  
at all photon energies $x$ for low temperatures and small degeneracy parameter $\eta$ 
as well as at $x \gtrsim \Theta$ for large $\Theta$ and $\eta$. 
The approximate expression  (\ref{eq:lx_nonrel}) used by \citet{Sampson:59} is also reasonably accurate for small $\Theta$ and $\eta$ for $x \gtrsim \Theta$, 
but becomes increasingly inaccurate for high $\Theta$ and $\eta$. 
We note that for large $\Theta$ and $\eta$ the solution of the integral equation  (\ref{eq:integr_mean}) gives negative $l_x$ at small $x$, which is unphysical; 
on the other hand, $l_x$ computed via Eq.~(\ref{eq:lx_rel})  is always positive.

\section{Rosseland mean opacity}
\label{sec:rosseland}

After finding the mean-free path $l_x$ as a solution of Eq. (\ref{eq:integr_mean}), we can compute the Rosseland mean opacity  as 
\be \label{eq:ross_opac} 
\kappaR = \frac{\sigmat \Ne}{ \rho } \frac{1 }{ \Lambda } , 
\ee   
where the Rosseland mean  free path 
(in units of $1/\sigmat \Ne$) is   
\be \label{eq:rosseland}
%\frac{}{\kappae} = 
%\strup
%\kappaR (\Theta,\eta_-) 
\Lambda (\Theta,\eta_-)  
= \displaystyle
\frac {\displaystyle \int_0^\infty  l_x \frac{\partial B_x}{\partial \Theta} \rmd x } 
{\displaystyle \int_0^\infty \frac{\partial B_x}{\partial  \Theta} \rmd x  }  
= 
%\left[ 
\frac{15}{4\pi^4} \int_0^\infty l_x  \frac{u^4 e^u}{(e^u-1)^2} \rmd u , 
%\right] ^{-1},
\ee
and $u=x/\Theta=h\nu/kT$ and $B_x=x^3 b_x$.  
The integrals over $x$ are taken over the energy range where $l_x$ is positive. 
We note that because of the high accuracy of  the approximation  (\ref{eq:lx_rel}), 
the Rosseland mean  can be also computed using explicit expression instead 
of solving integral equation (\ref{eq:integr_mean}),  giving typically the relative accuracy of better that $10^{-4}$. 
This approximation also allows us to easily find the photon mean free path when additionally 
true absorption needs to be accounted for:  $1/l_x \approx \alpha(x)+ r_0(x)-r_1(x)$, 
here $\alpha(x)$ is the standard absorption coefficient in  units $\sigmat \Ne$.

\begin{figure}
%\plotone{fig2.eps}
\centerline{\epsfig{file=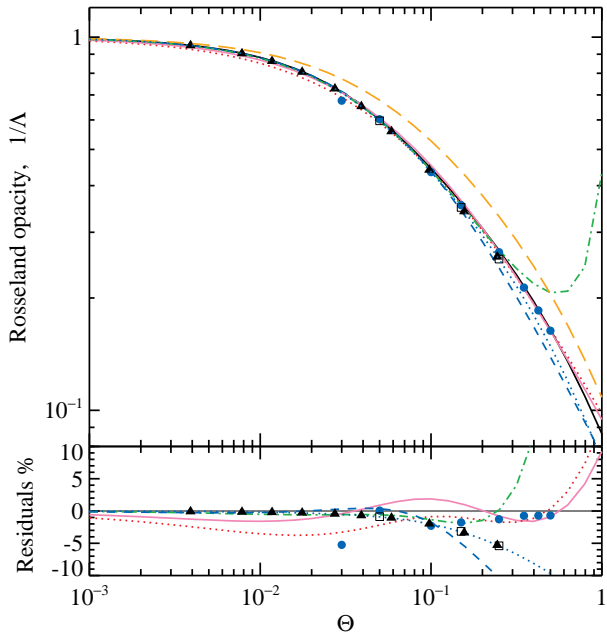,width=8.cm}} 
\caption{Rosseland mean opacity  as a function temperature for non-degenerate gas. 
Black solid curve represent the result of our exact calculations. 
The dotted blue curves give the Rosseland mean computed with the help of approximation  (\ref{eq:lx_nonrel}) for the mean free path. 
The  blue circles give the numerical results of  \citet{BY76}, the black triangles are the results from \citet{Sampson:59}, 
and the open squares are from \citet{Chin65}. 
The dotted red curve is the Paczy\'nski approximation (\ref{eq:pacz}), 
which underestimates exact results by 2--3 \%. 
The solid pink curve is the best fit in the temperature range 2--300 keV  using  function (\ref{eq:ourappr}) 
 with parameters $T_{0}=41.5$~keV and $\alpha_0=0.9$ which is accurate within 2\% in that range. 
The dashed blue curves is the same approximation in the temperature range 2--40 keV 
 with parameters $T_{0}=39.4$~keV and $\alpha_0=0.976$, which is accurate to 0.7\%. 
The dot-dashed green curve is the approximation (\ref{eq:wzw78}), which 
is accurate within 3\% in the temperature range of 1--150 keV and rapidly diverges at higher temperatures.  
The  long-dashed brown curve represent the flux mean opacity in the free-streaming limit (\ref{eq:fluxmean}) for the blackbody spectrum.
The bottom panel presents the residuals in per cent relative to our exact calculations. } 
\label{fig:opacity}
\end{figure}

\begin{figure}
%\plotone{fig3.eps}
\centerline{\epsfig{file=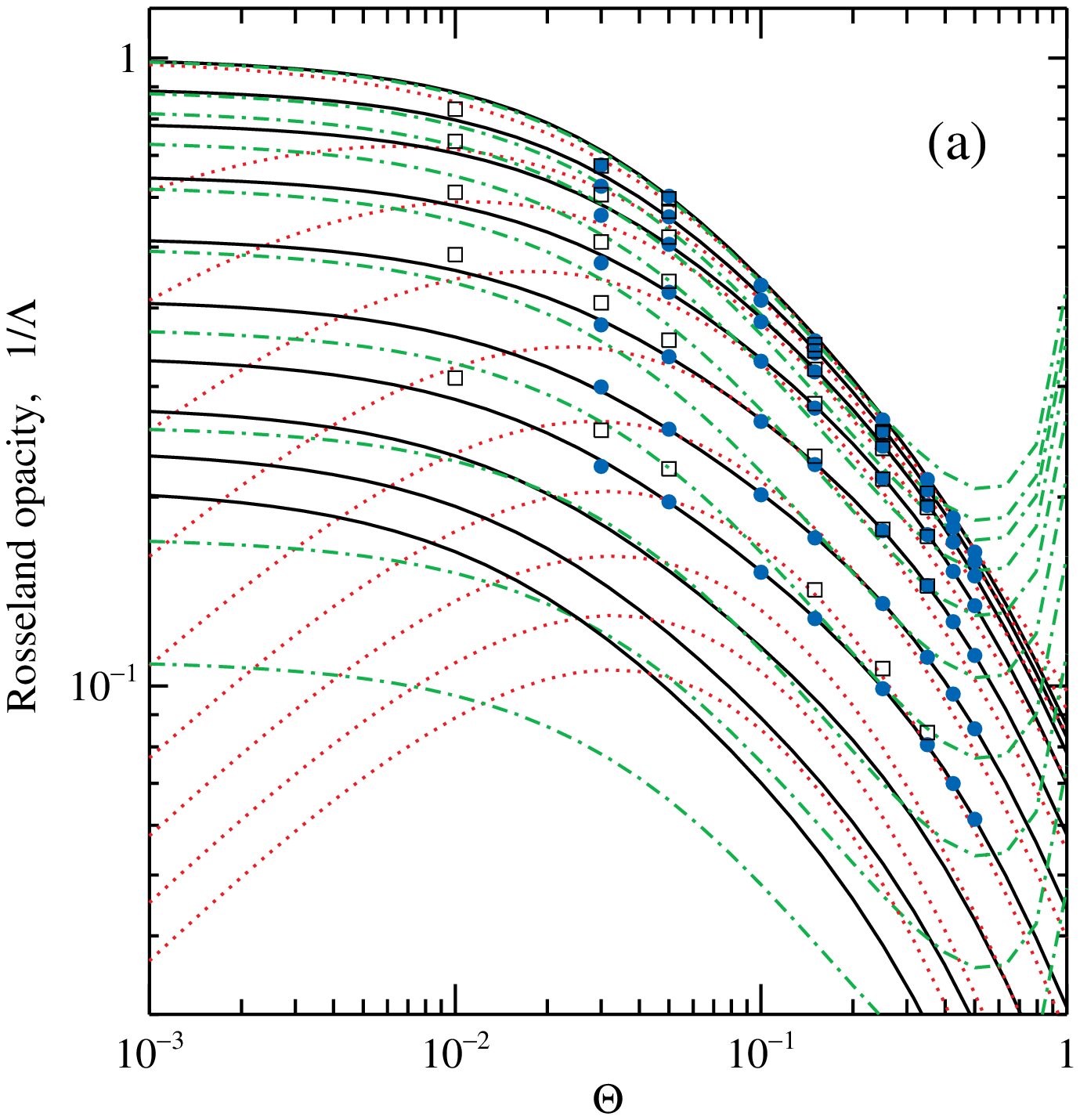,width=8cm} \hspace{1cm}
\epsfig{file=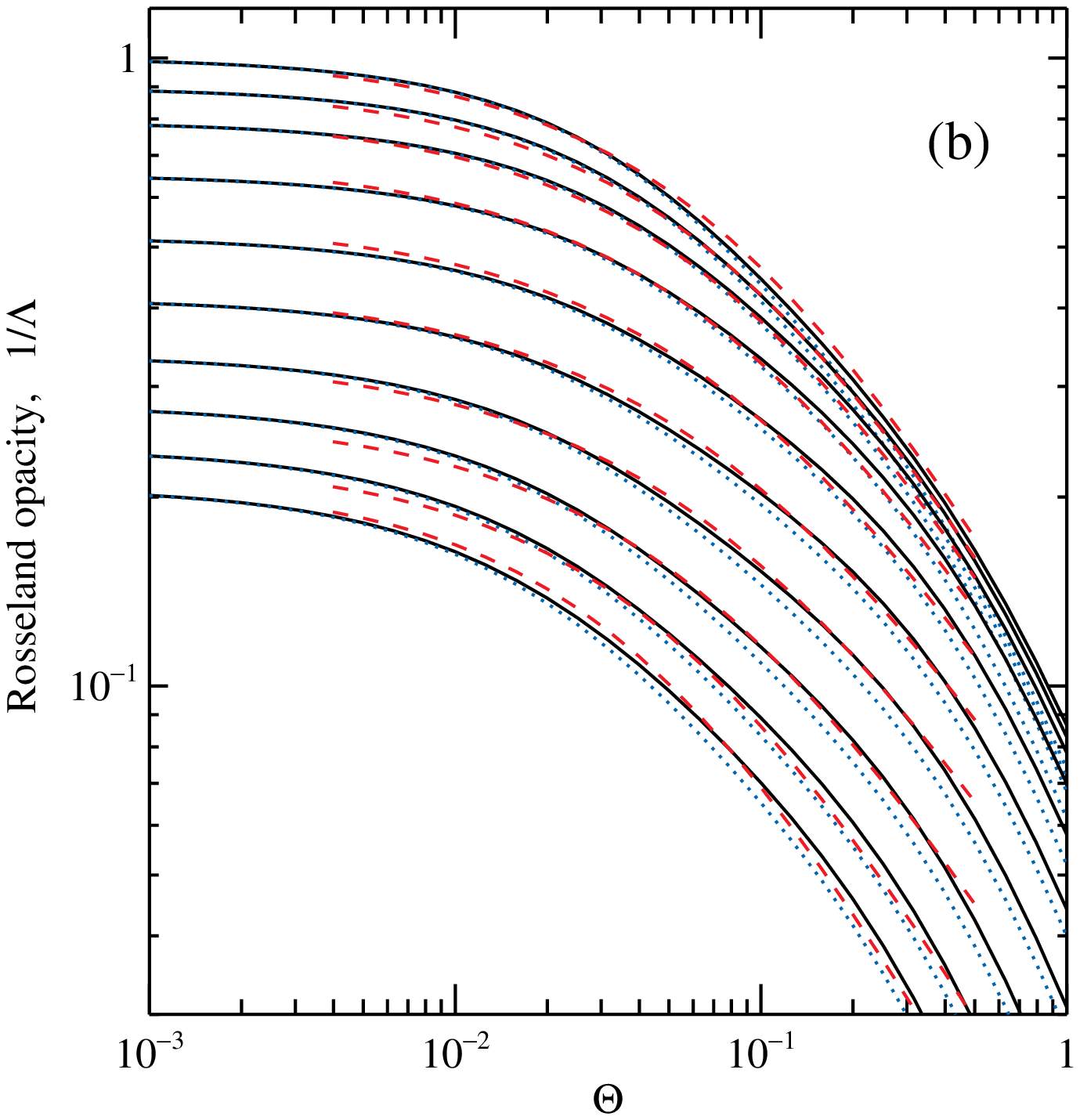,width=8cm} } 
\caption{Rosseland mean opacity  as a function of temperature for various values of the degeneracy parameter $\eta$. 
The solid black curves represent the result of our exact calculations (the flux mean opacity is equal to the Rosseland mean within $10^{-4}$).    
The top curve corresponds to $\eta=-10$, at the following curves $\eta$ varies from $-1$ to $7$.  
%The curves from top to bottom correspond to  $\eta=-10$ and then $-1,0,...,7$. 
(a)  The  blue circles give the numerical results of  \citet{BY76}  for $\eta=-10,-1,..,4$ 
and the  black open squares are from \citet{Chin65} for $\eta=-\infty,-1,0,1,2,4$ (for 5 values of $\Theta$, except for $\eta=-\infty$ for which 
the opacity is given for only three $\Theta=0.05, 0.15, 0.25$). 
The  dot-dashed green curves are the approximation (\ref{eq:wzw78}) 
and dotted red  curves give Paczy\'nski's approximation (\ref{eq:pacz}). 
(b) The dotted blue curves are the Rosseland mean computed using approximation  (\ref{eq:lx_nonrel}) for the mean free path and
the dashed red  curves correspond to our approximate  expression (\ref{eq:ourappr_deg}) for the 2--300 keV range.  
} 
\label{fig:opacity_degen}
\end{figure}

The results of calculations for $\Lambda$ in a broad range of temperatures 
and electron degeneracies $\eta_-$ are presented in Figs~\ref{fig:opacity} and \ref{fig:opacity_degen}. 
We present the results taking opacity by electrons only as was done also by \citet{Sampson:59} and \citet{BY76}, because 
at low temperatures or high degeneracy there are no pairs. At low degeneracies and high temperatures $\Theta>-1/\eta$, 
on the other hand, the number of positrons exceeds the number of electrons, because $\eta_+=-2/\Theta-\eta_->\eta_-$, which is unphysical.  
In the following we will replace $\eta_-$ by $\eta$.
%In Fig.~\ref{fig:opacity} we show the results of calculation of the Rosseland mean for non-degenerate electron gas ($\eta=-10$).
%\footnote{We note that these results become inconsistent 
%above temperature $\Theta=-1/\eta$, where photon-photon pair production will produce enough pairs 
%to that they will dominate the opacity. Formally, if we fix $\eta_-<0$, then at $\Theta>-1/\eta$  }
The results computed by \citet{Sampson:59} and \citet{BY76} are shown by triangles and circles, respectively, while our results by black solid curves. 
Results of \citet{Sampson:59} are accurate to better than 1\% up to about 25 keV, after that they start to deviate significantly.   
This is a direct consequence of his usage of approximation (\ref{eq:lx_nonrel}) for the mean free path, which is supported by 
our calculations in the same approximation (see dotted blue curves in Figs.~\ref{fig:opacity} and \ref{fig:opacity_degen}b, and the 
residuals in the bottom panel of  Fig.~\ref{fig:opacity}).  
We see that this approximation systematically underestimates the opacity at high temperatures. 
We note here that the opacity computed by \citet{Chin65} for degenerate electrons and still reprinted in the textbooks \citep{Weiss04} 
is systematically too large by up to 13\% (see black squares in Fig.~\ref{fig:opacity_degen}a); 
a rather good agreement at high temperatures results from a fortuitous cancellation of an error and his usage of approximation (\ref{eq:lx_nonrel}) \citep{BY76}.

On the other hand, results of \citet{BY76} are within 2\% from ours above 25 keV ($\Theta>0.05$), but at  $\Theta=0.03$  they underestimate the opacity by as much as 6\%. 
The situation becomes worse if we use the analytical approximations of \citet{BY76}  at lower temperatures, where the opacity would be systematically 
underestimated by up to 13\%.  

Calculations of \citet{BY76} gave rise to at least two different analytical formulae for the  
Rosseland mean opacity. \citet{WZW78} separated dependencies on   $\Theta$ and $\eta$: 
\be 
\label{eq:wzw78}
\Lambda_{\rm W78}  (\Theta,\eta) =   f_{\Theta} f_{\eta} ,  
\ee
where 
\be \label{eq:wzw78a}
%f_{\theta}  =  1+ 0.0275 \theta - 4.88\times 10^{-5} \theta^2 \ \theta< 200 \mbox{keV}, \\
f_{\Theta}  =  1+ 14.1 \Theta - 12.7 \Theta^2 \quad (\mbox{for } \ \Theta< 0.4) ,\qquad
f_{\eta}   =  1 + \exp(0.522\eta - 1.563). 
\ee
%and $\theta$ is the electron temperature in keV. 
Expressions (\ref{eq:wzw78})--(\ref{eq:wzw78a}) were claimed to be better than 10\% accurate  
in a wide range of  degeneracy parameters and temperatures 
($-\infty < \eta \lesssim 4$, $0.04<\Theta<0.4$). 
This approximation is used in the codes developed for simulation of the 
stellar evolution and explosions, including  X-ray bursts \citep{WHW02,Woosley04}.
We see (Fig.~\ref{fig:opacity_degen}a) it diverges above 150 keV for any $\eta$. 
Because the dependences on $T$ and $\eta$ are separated, 
the temperature range of applicability of this approximation becomes smaller for large $\eta$. 
For $\eta=4$ deviations from the exact values reach 50\% in the middle of the temperature range where 
the approximation suppose to work.  
 
A different approximation that is widely used in theory of X-ray bursts was given by  \citet{Pacz:83}: 
\be 
\label{eq:pacz}
\Lambda_{\rm P83}  (\Theta,\eta)  =  
\left[ 1+\left(\frac{kT}{38.8\, {\rm keV}} \right)^{0.86} \right] 
%\left[ 1+ 13.1\ \Theta ^{0.86} \right]^{-1}  
\left[ 1+2.7 \times 10^{11}\ \rho\ T^{-2} \right] 
%\left[ 1+7.7 \times 10^{-9}\ \rho\ \Theta^{-2} \right]^{-1}  
\ee 
for $\mue=2$.  We see from Figs~\ref{fig:opacity} and \ref{fig:opacity_degen}a that Paczy\'nski's 
approximation is rather good for small $\eta$. At large $\eta$ it becomes highly inaccurate at low temperatures.

\floattable 
\begin{deluxetable}{CCDCD}
\tablecaption{Coefficients of the approximate formulae (\ref{eq:ourappr}) and (\ref{eq:ourappr_deg}). \label{tab:coeff}}
\tablehead{
 \colhead{} & \multicolumn{6}{c} {Energy interval of the fit}    \\
 \colhead{Coefficient} &  \multicolumn3c{2--40 keV} &  \multicolumn3c{2--300 keV}   
}
%\colnumbers
\decimals
\startdata
 T_{0}    & $39.4$  & 43.4 & $41.5$ & 43.3   \\
 \alpha_0 & $0.976$  & 0.902 & $0.90$ &  0.885   \\
 c_{01}  &  &  0.777 & & 0.682 \\  
 c_{02}  &  & -0.0509 & &  -0.0454  \\  
 c_{11}  &  & 0.25 & & 0.24 \\  
 c_{12}  &  & -0.0045  & & 0.0043 \\  
 c_{21}  &  & 0.0264  & & 0.050 \\  
 c_{22}  & & -0.0033 &    & -0.0067 \\  
 c_{31}  & & 0.0046  &  & -0.037 \\  
 c_{32}  & & -0.0009 & & 0.0031 \\
\enddata
\end{deluxetable}

For the approximation of the Rosseland mean opacity, we propose to use 
a combination of the forms proposed by \citet{BY76} and \citet{Pacz:83}: 
\be \label{eq:ourappr_deg}
\Lambda_{\rm app}  (\Theta,\eta)   =   f_1 (\eta) \left[ 1+  (T/T_{\rm br})^{\alpha} \right] ,
\ee 
where 
\bea 
T_{\rm br}& =& T_{0} f_2 (\eta) ,  \\
\alpha& =&  \alpha_0 f_3 (\eta) , \\
f_i(\eta)& =& 1+c_{i1}\xi+c_{i2}\xi^2, \quad i=1,2,3, \\
\xi & =&  \exp(c_{01}  \eta +c_{02}\eta^2) . 
\eea 
The coefficients are given in Table~\ref{tab:coeff} for two fitting intervals 2--40  and 2--300 keV. 
This form approximates the opacity to better than 4\% and 6.5\% over the whole range of degeneracy parameter $-10,...,7$   in the temperature ranges 2--40 and 2--300 keV, respectively (see Fig.~\ref{fig:opacity_degen}b). 
In case of a non-degenerate gas, $\eta\rightarrow -\infty$, the expressions simplify as all $f_i=1$: 
\be \label{eq:ourappr}
\Lambda_{\rm app}  (\Theta)  =  1+  (T/T_{0})^{\alpha_0}   ,
\ee 
with the best-fit parameters given in Table~\ref{tab:coeff}. The goodness of the fits is demonstrated in Fig. \ref{fig:opacity}. 
%and the best description of the opacity in the 2--40 keV range is given by $T_{0}=39.4$ keV and $\alpha:0=0.976$ (blue dashed curve in Fig. \ref{fig:opacity}). 
%In a broader energy interval  of 2--300 keV, the best fit parameters are $T_{0}=41.5$ keV and $\alpha_0=0.9$ (pink solid curve in Fig. \ref{fig:opacity}); 
We note, however, that  physically realistic parameters should correspond to $N_->N_+$, which demands $\Theta<-1/\eta$, i.e. for $\eta=-10$ the temperature is limited by 50 keV.

\section{Radiative acceleration and the flux mean opacity} 
\label{sec:radpr}

For computation of the radiation  force on the medium, we need to construct the first moment of the RKE. 
Multiplying RKE (\ref{eq:rte2}) by $\vecx$ and integrating over $\rmd \vecx$, we get
\be \label{eq:tensor}
%\lefteqn{ 
%\frac{\vnabla\cdot T  }{\sigmat \: \Ne}  \!= \!
- \vnablat\cdot \tens{T}   \!= \! 
   \int \!\!\frac{\rmd \vecx}{x}\!\! \int\!\! \frac{\rmd \vecx_1}{x_1}  R(x_1,x,\mu) 
\noccx(\vecx)  [1+\noccx(\vecx_1)]
\left(  \vecx - \vecx_1 \right) , 
\ee
where we changed the variables $\vecx\leftrightarrow\vecx_1$ in 
the second half of the equation and introduced the (dimensionless) 
radiation pressure tensor 
\be \label{eq:press1}
\tens{T} =  \int \  \vecx\ \vecx\ \noccx(\vecx)\ \frac{\rmd \vecx}{x} . 
%T^{ij} =\frac{1}{c}  \int x \omega^i \omega^j \noccx(\vecx) \rmd \vecx
%T = %\frac{1}{c}  
%\int \fourx \fourx \noccx(\vecx) \frac{\rmd \vecx}{x}. 
\ee
This tensor is related to  the ordinary radiation pressure tensor as 
\be
\tens{P} =2\frac{\me c^2}{\lambdac^3} \tens{T} . 
\ee

Let us represent the gradient of the pressure tensor as a sum of  the linear and quadratic  in $\noccx$ terms: 
\be \label{eq:tensor_sum}
 \vnablat\cdot \tens{T} =    \vnablat\cdot \tens{T}_0   +  \vnablat\cdot \tens{T}_1 .   
\ee
Integrating over angles $\rmd^2\omega_1$, the first term becomes 
\be \label{eq:tensor2}
%\frac{\vnabla\cdot T  }{\sigmat \: \Ne}  \! & =&  \!
 - \vnablat\cdot \tens{T}_0   =
 \int \frac{\rmd \vecx}{x}\ \noccx(\vecx)\  4 \pi \vomega 
\int_0^\infty x_1\ \rmd x_1  
\left[   x R_0(x_1,x) - x_1 R_1(x_1,x)   \right] 
= 
%\int \rmd \vecx \ \noccx(\vecx)\  \vomega  
 4\pi \int  x^3 \rmd x \  \vech_x    \ [s_0(x)-s_1(x)]
%\ \overline{(x-x_1\mu)}\ \overline{s_0}(x)  ,
%\nonumber \\ &=&   
\ee
where 
\be
 \vech_x = \frac{1}{4\pi}  \int \  \vomega\ \noccx(\vecx)\  \rmd \vomega
\ee 
is the first moment of $n$ and 
%the flux 
%\be 
%\vech = \int \  \vecx\ \noccx(\vecx)\ {\rmd \vecx}  . 
%\ee
%and 
%$\overline{(x-x_1\mu)}\, \overline{s_0}(x)\Ne\sigmat\me c$ 
$x [s_0(x)-s_1(x)]\Ne\sigmat\me c$ 
is the momentum transfer  per units length of photon propagation averaged over 
electron distribution and scattered photon directions ignoring induced scattering \citep{NP94,PV10,PSS83}. 
Note the radiation flux in these notations is 
\be 
\vecF \propto 
%=  \frac{8\pi \me c^3}{\lambdac^3} 
\int  x^3 \rmd x \  \vech_x \  . 
\ee
Thus the flux mean opacity (in units of $\Ne\sigmat/\rho$) in the free streaming limit (ignoring induced scattering) is 
\be \label{eq:fluxmean}
\kappa_{\rm F} = \frac{\int  x^3 \rmd x \  h_x    \ [s_0(x)-s_1(x)]}{ \int  x^3 \rmd x \  h_x } .
\ee
For the radiation spectrum close to the diluted blackbody, i.e. $h_x\propto b_x$, 
the flux mean opacity for the case of non-degenerate electrons  is shown in  Fig.~\ref{fig:opacity}. 
Eq.~(\ref{eq:ourappr}) with parameters $T_0=58.5$~keV  and $\alpha_0=0.913$ provides a good  approximation with better than 0.8\% accuracy to the corresponding mean free path 
in the 2--300 keV range. 
 
In the  diffusion approximation (\ref{eq:n_diffuse}), we substitute $\vech_x=- \frac{1}{3}l_x \vnablat b_x$. 
%and substitute it to equation (\ref{eq:tensor}). 
%\be\label{eq:n_diffuse}
%\noccx(\vecx) = b_x + {l_x} \omega\cdot \vnablat b_x ,
%\ee
The effect of the induced scattering on the radiation pressure force in the  diffusion approximation can be computed by 
substituting equation (\ref{eq:n_diffuse}) to Eq.~(\ref{eq:tensor}). 
%Projecting this on the direction of the temperature gradient, 
The non-linear term becomes: 
\be
% \frac{ \rmd {T}_1 }{\rmd \taut}  
- \vnablat\cdot \tens{T}_1 = 
   \int  x \rmd x   \int   x_1 \rmd x_1  
 \int \rmd^2\omega    \int \rmd^2\omega_1
 \left(   b_x + \eta l_x \partialt b_{x}\right)  \left(   b_{x_1} + \eta_1 l_{x_1} \partialt b_{x_1} \right)   
 R(x_1,x,\mu)  \left(  x\vomega - x_1\vomega_1 \right) , 
\ee
where  $\vomega=(\sqrt{1-\eta^2}\cos\phi, \sqrt{1-\eta^2}\sin\phi,\eta)$ 
and $\vomega_1=(\sqrt{1-\eta_1^2}\cos\phi_1, \sqrt{1-\eta_1^2}\sin\phi_1,\eta_1)$  with the $z$-axis chosen against the temperature gradient  and 
$\partialt b_{x} \equiv  |\vnablat b_x| $. 
Taking the angular integrals, for the magnitude of the radiation pressure force  we  get 
%\bea
% \frac{ \rmd {T}_1 }{\rmd \taut}  \! &=& \!  16\pi^2 
%   \int \!\! x \rmd x \!\! \int\!\!   x_1 \rmd x_1  
% \left[   R_0(x_1,x)  \left( x  j_{x_1} h_x  -x_1  j_{x} h_{x_1}\right)  \right. \nonumber \\
% &+& \left. R_1(x_1,x)   \left( x j_{x} h_{x_1}   -x_1   j_{x_1} h_x \right)  
% \right]   .
% \eea 
%Now if we assume that the flux and the mean intensity have the same energy 
%distribution with $h_x=\alpha j_x$, we get 
%\be
% \frac{ \rmd {T}_1 }{\rmd \taut}   \! = \!  16\pi^2 \alpha 
%   \int \!\!  j_{x} x  \rmd x \!\! \int\!\!   j_{x_1} x_1 \rmd x_1  
%   (x-x_1)  \left[   R_0(x_1,x)   + R_1(x_1,x)    \right]   .
%\ee 
%In the  diffusion approximation, we substitute here equation (\ref{eq:hx_diff}) to get
\be
%\partialt {T}_1   
 \! \!- |  \vnablat\cdot \tens{T}_1 |
=   \frac{16\pi^2}{3}   \! \!
   \int  \! \! x \rmd x \! \! \int   \! \!  x_1 \rmd x_1  
 \left[   R_0(x_1,x)  \left( b_{x_1} x l_x  \partialt  b_{x}   - b_{x}  x_1 l_{x_1} \partialt b_{x_1} \right)  
+ R_1(x_1,x)   \left( x b_{x} l_{x_1} \partialt b_{x_1}    -x_1  b_{x_1} l_{x} \partialt b_{x} \right)  
 \right]   .
 \ee
Making variable change $x\leftrightarrow x_1$ in the terms containing $\partialt b_{x_1}$, we get 
 \be
- |  \vnablat\cdot \tens{T}_1 |  =  \frac{16\pi^2}{3}  
   \int \!\! x \rmd x\  l_x  \partialt b_{x}   \!\! \int\!\!   x_1 \rmd x_1   b_{x_1} 
\left\{  \left[ x  R_0(x_1, x)-  x_1  R_1(x_1,x)     \right] 
 -\left[x R_0(x, x_1)  -x_1 R_1(x,x_1)    \right] \right\} . 
\ee
%\be
% \frac{ \rmd {T}_1 }{\rmd \taut}   \! = \!  16\pi^2 \alpha male
%   \int \!\!  j_{x} x  \rmd x \!\! \int\!\!   j_{x_1} x_1 \rmd x_1  
%   (x-x_1)  \left[   R_0(x_1,x)   + R_1(x_1,x)    \right]   .
%\ee 
The total pressure gradient becomes  
\bea
%\partialt  {T} 
  \! \!\! \! \! \! - |  \vnablat\cdot \tens{T} |
& =&   \frac{16\pi^2}{3}  
   \int x \rmd x\  l_x  \partialt b_{x}   \int   x_1 \rmd x_1   
 \left\{   (1+ b_{x_1}) \left[ x  R_0(x_1, x)-  x_1  R_1(x_1,x)   \right]  
  - b_{x_1} \left[x R_0(x, x_1)  -x_1 R_1(x,x_1)    \right] \right\}  \nonumber \\
 &=  &  
  \frac{16\pi^2}{3}  \! \! \!   \int  \! \!  x \rmd x\  l_x  \partialt  b_{x}   \int  \! \!  x_1 \rmd x_1   
   \frac{1-e^{-x/\Theta}}{1-e^{-x_1/\Theta}}   
 \left[ x  R_0(x_1, x)-  x_1  R_1(x_1,x)   \right]  =
 \frac{4\pi}{3} \! \!     \int  \! \!  x^3 \rmd x\  l_x  \partialt  b_{x} \left[ r_0(x) -r_1(x) \right]   .
\eea
%\bea
% \frac{ \rmd {T} }{\rmd \taut}  \! &=& \!  4\pi 
%   \int \!\!  F_{x} \frac{\rmd x}{x} \!\! \int\!\! \frac{x_1^2}{x} \frac{\rmd x_1}{x_1}  
% \left\{   x R_0(x_1,x) - x_1 R_1(x_1,x) \right. \nonumber \\
% &+& \left.  j_{x_1}   (x-x_1)  \left[   R_0(x_1,x)   + R_1(x_1,x)    \right]  \right\} ,  
%\eea
%where $F_x=\alpha x^3 j_x$ is the energy flux. 
%This result implies that contribution of the induced scattering 
%to the total radiation pressure force does not depend on the 
%amount of radiation beaming (determined by parameter $\alpha$), but 
%just by the occupation number. The closer radiation field 
%is to the blackbody, the higher is the contribution of induced scattering.  
Thus the flux mean opacity  is then 
\be
\kappa_{\rm F}     = 
\frac{ \displaystyle \int_0^\infty l_x \   [ r_0(x)-r_1(x)] \frac{\partial B_x}{\partial \Theta} \rmd x   }
{\displaystyle \int_0^\infty  l_x \frac{\partial B_x}{\partial \Theta} \rmd x  }   . 
\ee
Because approximation (\ref{eq:lx_rel}), i.e. $1/l_x \approx r_0(x)-r_1(x)$, is very accurate 
in the region of the photon energies $x\sim\Theta$ contributing to the integral, 
the flux mean opacity in the diffusion approximation turned out to be nearly identical (with the relative difference less than $10^{-4}$) 
to the Rosseland mean in the full range of considered temperatures and degeneracies. 
The flux mean opacity thus can  be approximated by simple analytical expressions (\ref{eq:ourappr_deg}) and  (\ref{eq:ourappr}),
which also describe well the radiative acceleration in the atmospheres of hot neutron stars 
obtained from the solution of the radiative transfer equation with the exact Compton redistribution function 
where diffusion approximation has not been used \citep{SPW12}.

%Our stellar atmosphere calculations \citep{SPW12}
%using exact RF showed that approximation (\ref{eq:pacz}) is not good in the outer cooler layers. 

\section{Summary}  
\label{sec:summ}

In this paper, we have critically evaluated results of previous works on the Rosseland mean opacity  for Compton scattering. 
In order to obtain  the photon mean free path as a function of photon energy 
we have solved relativistic kinetic equation describing photon interactions via Compton scattering with the possibly degenerate electron gas. 
We demonstrated that the mean free path  can be also accurately evaluated using explicit approximate formulae, 
which can also be used for calculations of the Rosseland mean opacity and 
can provide a simple way for accounting for the true absorption.

We have computed the Rosseland and the flux mean opacities (which are nearly identical in the diffusion approximation) 
in a broad range of temperatures and electron degeneracy parameter. 
We compared our results to the previous calculations and found significant difference in the low-temperature regime. 
We have also presented useful analytical expressions that approximate well the numerical results.

\acknowledgments
The author thanks Valery Suleimanov, Dmitri Nagirner and Dmitry Yakovlev for useful discussions.
This work was supported  by the Foundations' Professor Pool, the Finnish Cultural Foundation and the Academy of Finland grant 268740.  
The author also acknowledges useful conversations with the members of the X-ray burst team 
of the International Space Science Institute (Bern, Switzerland).

\appendix

\section{Redistribution functions}
\label{sec:appa}

In case of non-degenerate electrons, the RF defined by Eq.~(\ref{eq:rf_gen}) 
has been studied  in details before \citep{AA81,PKB86,NP94,PV10}. 
This RF can be simplified to the one-dimentional integral over the electron energy. 
For degenerate electrons, it turned out that the derivation is  identical 
and also in this case, the RF can be presented in terms of one integral 
that can be taken numerically.  

For the isotropic electron distribution expression (\ref{eq:rf_gen}) for the RF can be simplified 
by taking the integral over $\vecp$ with the help of  the 3-dimensional delta-function
and using the identity $\delta(\gamma_{1}+x_1-\gamma-x)=\gamma 
\delta \left( \fourx_1 \cdot \fourp_1- \fourx \cdot (\fourp_1+ \fourx_1) \right)$: 
\be \label{eq:rf_spe}
R_\pm(x,x_1,\mu) =  \frac{3}{16\pi}   \frac{2}{\lambdac^3 N_\pm} \!\!  
\int \!\!  \frac{\rmd \vecp_1}{\gamma_1} \noccpm(\gamma_1) [1-\noccpm(\gamma)] 
 F \delta(\Gamma) ,
\ee
where 
%we dropped subscript 1 with the electron quantities and 
\bea
\gamma & =&  \gamma_1+x_1-x, \\
\Gamma& =& \gamma_1(x_1-x) - p_1 (x_1 \vomega_1 -x \vomega ) \cdot \vOmega_1 - q , \\ %\quad 
q &=&\vecx\cdot\vecx_1 = xx_1(1-\mu) . 
\eea
The RF depends also implicitly on the electron temperature $\Theta$ and degeneracy 
parameter $\eta_\pm$. 
To integrate over angles in Eq.~(\ref{eq:rf_spe}) we 
follow the recipe proposed by \citet{AA81} (see also \citealt{PKB86,PV10}), 
choosing the polar axis along the direction of the transferred momentum
\be 
\vn \equiv \left( x_1 \vomega_1 -x \vomega\right) / Q ,
%\frac{x_1 \vomega_1 -x \vomega}{Q} ,
\ee
where 
\be \label{eq:Qcap}
Q^2=(x_1 \vomega_1 -x \vomega)^2= 
%x^2+x_1^2-2xx_1\eta = 
(x-x_1)^2+2q. 
\ee 
Thus the integration variables become $\cos\alpha = \vn\cdot\vOmega_1$ and the corresponding azimuth $\Phi$. 
The RF (\ref{eq:rf_spe}) then can be written as
\be\label{eq:red_sim}
R_\pm(x,x_1,\mu)  =  \frac{3}{16\ \pi}  \frac{2}{\lambdac^3 N_\pm}  
 \!\int\limits_{1}^{\infty} \!\! \, \noccpm(\gamma_1) [1-\noccpm(\gamma)] p_1 \rmd \gamma_1 
 \int \limits_{-1}^{1} \!\! \, \delta(\Gamma)\ \rmd\cos\alpha  %\nonumber \\  && 
\!\! \int \limits_{0}^{2\pi} \!\! \, F  \rmd\Phi ,	
\ee
where now
\be 
\Gamma= \gamma_1(x_1-x) - q -  p_1  Q \cos\alpha .
\ee 
Integrating   over $\cos\alpha$ using  the delta-function we get 
\be \label{eq:red_phi}
R_\pm(x,x_1,\mu)   =    \frac{3}{8} \frac{2}{\lambdac^3 N_\pm}   
\int_{ \gamma_{*}}^{\infty} \!\!
\noccpm(\gamma_1) [1-\noccpm(\gamma)]\  R(x,x_1,\mu,\gamma_1)\ \rmd \gamma_1 ,
\ee
with integration over the electron distribution done numerically. 
Here we introduced the RF for monoenergetic electrons 
\be\label{eq:rfmono_def}
R(x,x_1,\mu,\gamma_1) = \frac{1}{Q} \frac{1}{2\pi} \int_{0}^{2\pi} \: F\ \rmd\Phi  . 
\ee
Function $F$  depends on $\xi_1=x(\gamma_1-p_1\zeta)$ and $\xi=\xi_1+q$, where 
\bea 
\zeta&=&\vomega\cdot\vOmega_1=\cos\alpha\cos\kappa+\sin\alpha\sin\kappa\cos\Phi, \\
\label{eq:costheta}
\cos\alpha&=&  \vn\cdot\vOmega_1 = \left[\gamma_1 (x_1-x) - q\right]/p_1 Q , \\
\cos\kappa&=&\vomega\cdot \vn=(x_1\mu-x)/Q, 
\eea
The condition $|\cos\alpha|\le1$, gives a constraint 
\be\label{eq:gammastar}
\gamma_1 \ge \gamma_* (x,x_1,\mu) = \left(x-x_1 +Q\sqrt{1+2/q} \right)/2 .
\ee
Integrating over azimuth $\Phi$ in Eq.~(\ref{eq:rfmono_def}) gives the exact analytical 
expression for the RF valid for any photon and electron energy \citep{BY76,AA81,PKB86,NP94,PV10}, 
which we use in our calculations: 
\be\label{eq:r0f}
R(x,x_1,\mu,\gamma_1) \! = \!
\frac{2}{Q} + \frac{q^2-2q-2}{q^2} \left( \frac{1}{a_-}  \! - \!\frac{1}{a_+} \right)
+ \frac{1}{q^2} \left( \frac{d_-}{a_-^3}  \! + \! \frac{d_+}{a_+^3} \right), 	
\ee
where 
\be
a_-^2  = (\gamma_1-x)^2 + \frac{1+\mu}{1-\mu}    , \quad 
a_+^2 = (\gamma_1+x_1)^2 + \frac{1+\mu}{1-\mu}  ,  \quad
d_{\pm}  =  \left( a_+^2 -  a_-^2 \pm Q^2\right)/2 . 
%\gamma(x+x_1) - x(x-x_1 \mu)  , 
%\quad  d_+ =   d_- + Q^2. 
\ee
The cancellations at small photon energies has been handled by \citet{NP93}. 
We note that the RF (\ref{eq:rfmono_def}) satisfies the detailed balance condition \citep{NP94}: 
\be\label{eq:rfmono_bal}
R(x,x_1,\mu,\gamma_1)  = R(x_1,x,\mu,\gamma_1+x_1-x).  
\ee

The RF (\ref{eq:rf_spe}) is related to the scattering kernel (8.13) in \citet{Pom73} as 
$R(x,x_1,\mu)= \sigma_s(x_1\rightarrow x,\mu)x_1/x$. 
The form given by Eq.~(\ref{eq:red_sim}) is equivalent to eq. (A4) in \citet{BY76}. 
The derived RF  for monoenergetic electrons (\ref{eq:r0f}) 
is equivalent to eq.  (A5) from \citet{BY76} and eq. (14) in \citet{AA81}. 
 
The angle-averaged RF functions (\ref{eq:RF_moment_0}) and (\ref{eq:RF_moment_1}), 
used in the calculations of the mean free path, 
can be expressed through the single integral over the electron and positron distributions 
\be\label{eq:RF_moments2}
R_{i}(x,x_1) =   \frac{3}{16}  \frac{2}{\lambdac^3\Ne}  \int_{ \gamma'_{\star}}^{\infty} 
 R_{i}(x,x_1,\gamma_1)\ \rmd \gamma_1 
%\nocce(\gamma_1) [1-\nocce(\gamma)] 
 \left\{ \noccm(\gamma_1) [1-\noccm(\gamma)] + \noccp(\gamma_1) [1-\noccp(\gamma)]  \right\} , \quad i=0,1 ,\nonumber
%\label{eq:RF_moments3}
%R_1(x_1,x) & = &   \frac{3}{16}  \int_{ \gamma'_{\star}}^{\infty} \: \fe (p)\  R^*(x,x_1,\gamma)\ \rmd \gamma, 
%R_1(x,x_1) &\! =\! &   \frac{3}{16} \frac{2}{\lambdac^3\Ne} \!\!   \int_{ \gamma'_{\star}}^{\infty} \!\!\nocce(\gamma_1) [1-\nocce(\gamma)]
%R_1(x,x_1,\gamma_1)\ \rmd \gamma_1, 
\ee
where 
\begin{equation}\label{eq13}
\gamma'_{\star}(x,x_1)\!\!  =\!\! 
%\begin{cases}
\left\{
\begin{array}{ll}
\displaystyle 
\frac{x - x_1}{2} + \frac{x+x_1}{2}\!\! \sqrt{1+\!\!\frac{1}{xx_1} }  & \mbox{if $|x-x_1| \ge 2xx_1$,}   \\
1 + \left(x - x_1 + |x - x_1| \right)/2 &  \mbox{if $|x-x_1| \le 2xx_1$.} 
%\end{cases}
\end{array} \right. 
\end{equation}
The explicit analytical expressions for the angle-integrated functions $R_0(x,x_1,\gamma_1)$ and $R_0-R_1$ under the integrals in Eqs. (\ref{eq:RF_moments2}) 
%and (\ref{eq:RF_moments3}) 
%(actually for functions $\overline{R}\equiv R_0$ and $\overline{R^*}\equiv R_0-R_1$)  
can be found in sections 8.1 and 8.2 of \citet{NP94}.

%\bibliographystyle{apj}
%\bibliography{xrb}

\end{document}